# Building User-defined Runtime Adaptation Routines for Stream Processing Applications

Gabriela Jacques-Silva[†], Buğra Gedik[†‡], Rohit Wagle[†],
Kun-Lung Wu[†], Vibhore Kumar[†]

[†]Thomas J. Watson Research Center, IBM Research, Hawthorne, NY 10532, USA
[‡]Computer Engineering Department, Bilkent University, Ankara 06800, Turkey
{g.jacques,rwagle,klwu,vibhorek}@us.ibm.com, bgedik@cs.bilkent.edu.tr

## ABSTRACT

Stream processing applications are deployed as continuous queries that run from the time of their submission until their cancellation. This deployment mode limits developers who need their applications to perform runtime adaptation, such as algorithmic adjustments, incremental job deployment, and application-specific failure recovery. Currently, developers do runtime adaptation by using external scripts and/or by inserting operators into the stream processing graph that are unrelated to the data processing logic. In this paper, we describe a component called *orchestrator* that allows users to write routines for automatically adapting the application to runtime conditions. Developers build an orchestrator by registering and handling events as well as specifying actuations. Events can be generated due to changes in the system state (e.g., application component failures), built-in system metrics (e.g., throughput of a connection), or custom application metrics (e.g., quality score). Once the orchestrator receives an event, users can take adaptation actions by using the orchestrator actuation APIs. We demonstrate the use of the orchestrator in IBM's System S in the context of three different applications, illustrating application adaptation to changes on the incoming data distribution, to application failures, and on-demand dynamic composition.

## 1. INTRODUCTION

Stream processing applications perform a sequence of transformations on live data streams. Developers build these applications by composing a *data flow graph*, where each vertex of the graph is an *operator* instance and each edge is a *stream* connection. An operator executes data transformations upon the arrival of a stream data item, referred to as a *tuple*, and sends the newly computed data item to its output streams. For achieving high-performance and scalability, the stream processing graph can execute in a distributed fashion over a set of hosts.

Developers deploy a streaming application by submitting the composed flow graph to the target stream processing infrastructure, which then continuously runs the application until it is explicitly cancelled. Multiple applications can be submitted to the infrastructure at different times. These applications can connect to each other at run-time to form time-evolving solutions.

Oftentimes streaming applications need to automatically adapt to runtime conditions. For instance, when the application is overloaded due to a transient high input data rate, it may need to temporarily apply load shedding policies to maintain answer timeliness [25]. As a second example, different streaming applications can have different requirements regarding fault tolerance [15, 16] (e.g., tuple loss tolerant, no tuple loss or duplication allowed). As a result, they need application-specific routines to coordinate failure recovery. As a third example, applications may also adapt due to events related to their own data processing semantics. For example, one may choose to deploy a low resource consumption streaming algorithm $A$ at first, but switch to a more resource hungry and more accurate streaming algorithm $B$ when a certain pattern is detected (such as low prediction accuracy).

Stream processing languages usually do not provide means to express runtime adaptation. This is because these languages are generally declarative, such as StreamSQL [17], CQL [5], StreamIt [21], and IBM's SPL [11]. As a result, developers focus on expressing data processing logic, but not orchestrating runtime adaptations.

A common approach to making application executions adaptable is to resort to external scripts that interact with the stream processing infrastructure. These scripts are generally hard to maintain and their logic have very low reusability among different applications. This is because different developers use different scripting languages, achieving the same task in many different ways (e.g., using widely different shell commands).

As an example, a script can use command line tooling provided by the streaming infrastructure to monitor health of application components and initiate application-specific failure recovery actions upon detection of failures. Examples of such actions include restarting application components, cleaning up residual data files, or interacting with external data stores to coordinate recovery.

Another approach to accomplishing adaptability is to develop application extensions that deal specifically with adaptation logic, leading to the coupling of the application control logic and the stream data processing logic. These application extensions are possible in languages and frameworks that allow user-defined functionality, such as SPL and the Microsoft .NET framework LINQ used in StreamInsight [2]. Unfortunately, such extensions limit the reusability of the applications, since the processing logic (often highly reusable across applications) is tied to the control logic (often varies across applications).

Figure 1 shows a simplified example based on a sentiment analysis application implemented in SPL in which the processing graph is extended to provide runtime adaptation. In this example, the application consumes a Twitter (http://www.twitter.com) feed ($op_1$) and categorizes each incoming tweet according to a sentiment ($op_3$). The tweet is then correlated to a set of possible causes for negative sentiment associated with products ($op_5$). The list of causes is computed offline on a large corpus and loaded by the streaming application to do online tweet categorization ($op_2$ and $op_4$). The results of the correlation are then aggregated ($op_6$) and sent to a display application ($op_7$). The aggregation stream is also used to trigger application adaptation ($s'$). The triggering condition (monitored by $op_8$) is that the number of tweets with negative sentiment associated with an unknown cause is high. When $op_9$ receives a tuple from $op_8$, $op_9$ calls an external script that invokes the cause recomputation on a larger and more recent corpus. The new set of causes is then automatically reloaded by operators $op_2$ and $op_4$. Because the control logic is embedded into the application graph, neither the data processing logic nor the adaptation logic can be reused by other applications.

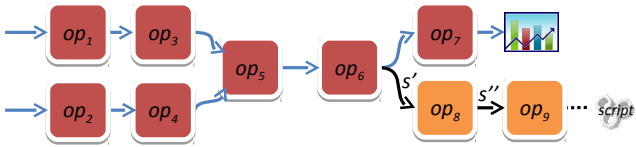

**Figure 1: Example application that includes extra operators to implement the adaptation logic. Operator $op_8$ detects the condition for adaptation and operator $op_9$ executes the actuation logic.**

To avoid the development of ad-hoc adaptation solutions for streaming applications and at the same time separate the control and data processing logic, we propose a new component for stream processing systems called *orchestrator*. An orchestrator is divided in two parts. The first part, called *orchestrator logic* (or ORCA logic), is put together by application developers by registering for runtime events of interest and specifying handlers that will be executed upon the delivery of these events. Such handlers often make use of actuation APIs to facilitate runtime adaptation. The second part, called *orchestrator service* (or ORCA service), is a runtime component that detects changes and delivers relevant events to the ORCA logic. The ORCA logic can further use this service to inspect the meta-data associated with the running application components to carry on specific tasks.

By running an orchestrator together with the application, developers can effectively make their application follow a specific management policy. The orchestrator also promotes code reusability by promoting separation of control and data processing code. This allows the application to be reused by other solutions that have different runtime adaptation policies. In addition, it provides a uniform way to implement event detection and actuation policies. This enables developers to share their adaptation solutions.

Developing an orchestrator for streaming applications has two main challenges. First, streaming applications can have distinct logical and physical representations. This is because the stream operator graph can be physically separated into different operating system processes and run on a distributed set of hosts. As a result, adaptation policies need to be able to understand and possibly influence the physical representation of the processing graph. Yet, the applications are developed using the logical representation of the graph. This implies that the mapping between the logical and the physical representation needs to be made available to the developers via the orchestrator.

The second challenge is that the orchestrator must be able to express interest on dynamic properties (such as state, health, and metrics) of runtime instances of application components, and receive relevant events with enough context information to associate them back to the application components without ambiguity. This requires an effective interface to register for events over these properties in the presence of hierarchical representation of application components (both at the logical and physical levels) and deliver these events with sufficient context information to facilitate effective actuation to implement autonomous behavior.

To tackle these challenges, our orchestrator design is based on three key concepts. The first is a *flexible event registration* scheme, where filters that use the properties defined over the logical application view are employed to express application interests with clarity. The second is the *event context*, which is associated with each event that the ORCA service delivers to the ORCA logic. The context contains physical and logical information about the application runtime slice associated with a given event type and instance. The third is an *in-memory stream graph representation* that has both logical and physical deployment information. This representation is maintained by the ORCA service and can be queried by the adaptation logic using an event context (e.g., which other operators are in the same operating system process as operator $x$?). This allows users to understand the relationship between the logical and physical mapping of the application to take appropriate adaptation actions.

We implemented an orchestration framework on top of System S [4, 27] – IBM's middleware for stream processing. Our current implementation is a C++ API, which, in the future, will become part of the SPL [11] language. We demonstrate the use of the orchestrator for building automatically managed streaming applications by showcasing the following three scenarios: (i) using application semantic-related events to detect changes on the incoming data and trigger an external model re-computation; (ii) using process failure events to trigger replica failover; and (iii) using events to start new streaming applications, which can consume data produced by other applications under execution. These use cases show how the orchestrator can effectively tackle real-world scenarios that require autonomic management.

The main contributions of this paper are (i) a framework based on event handling for building adaptive streaming applications, which allows users to separate the control and data processing logic of an application; (ii) the representation of event context and the meta-data of a stream graph, which enables developers to disambiguate logical and physical views of an application; (iii) a framework for managing a set of streaming applications with dependency relations, which includes automatic submission of dependent appli-cations and automatic cancellation of unused applications; and (iv) the demonstration of scenarios associated with real streaming applications that require orchestration solutions. We believe that these concepts and use cases are generic and can be applied to other stream processing platforms.

## 2. OVERVIEW OF IBM SYSTEM S

System S is IBM's middleware for developing high-performance and highly scalable stream processing applications. System S can be divided into two main parts: the Streams Processing Language (SPL) and its runtime components.

### 2.1 Streams Processing Language

SPL [11] allows the composition of streaming applications by assembling operators and expressing their stream interconnections. In SPL, operators can implement any logic (e.g., filtering, aggregation, image processing) and be arbitrarily interconnected. The language allows developers to define *composite operators*. A composite operator is a logically related sub-graph that can be reused to assemble more complex graphs. Composite operators are important for application modularization, similar to methods and classes in object-oriented programming. Figure 2 shows an SPL application that uses a composite operator ($composite_1$) to express a split and merge operation formed by four operators ($op_{3-6}$). The application instantiates the composite operator twice to process data from both $op_1$ ($composite_1'$) and $op_2$ ($composite_1''$).

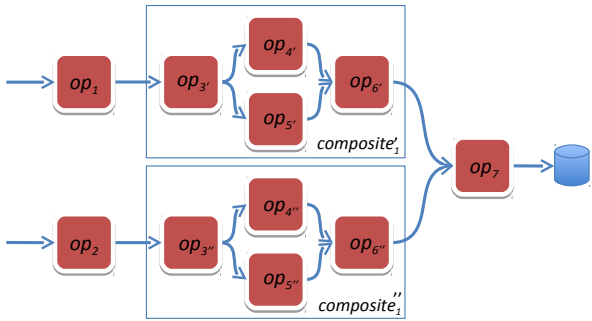

Figure 2: Stream graph generated by an SPL program that reuses a split and merge composite operator to build an application.

To execute an application, the SPL compiler places operators into *processing elements* (PEs), which are runtime containers for one or more operators. During execution, each PE maps to an operating system process, which can execute in any host available to the stream processing infrastructure. The compiler partitions operators into PEs based on performance measurements and following *partition* constraints informed by the developers (e.g., $op_{4'}$ and $op_{6'}$ should run in the same PE) [18]. During runtime, PEs are distributed over hosts according to *host placement* constraints informed by developers (e.g., PEs 1 and 3 cannot run on the same host) as well as the resource availability of hosts and load balance. The SPL compiler can group operators that belong to different composites into the same PE. This means that the physical streaming graph layout does not reflect the fact that some operators are logically grouped. Figure 3 shows a possible physical mapping of the application illustrated in Figure 2. Note that operators in the same composite are split into two different PEs ($op_{3'-6'}$ in PEs 1-2). Operators in different composites instances ($composite_1'$ and $composite_1''$) are placed in the same PE ($op_{4'-6'}$ and $op_{4''-6''}$ in PE 2).

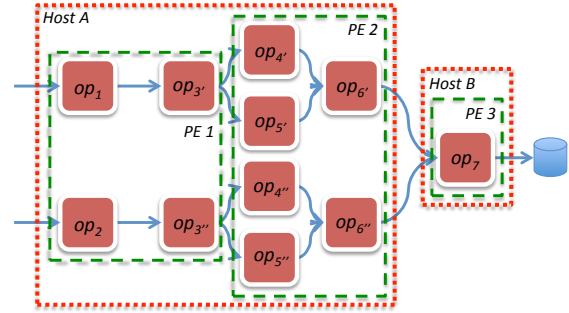

Figure 3: Possible physical layout of the application in Figure 2. Application is partitioned into three PEs and placed on two different hosts.

When the SPL compiler builds an application, it generates C++ code for each used operator and a file with the application description called ADL. The ADL is an XML description that includes the name of each operator in the graph, their interconnections, their composite containment relationship, their PE partitioning, and the PE's host placement constraints. Both the System S runtime and its visualization tools use the ADL for tasks such as starting the application and reporting runtime information to the users.

Another feature of SPL is runtime metrics. These metrics are counters updated during application execution and can be read externally by users to inspect runtime statistics. SPL offers both built-in and custom metrics. Built-in metrics are counters that maintain information that is common to all operators and PEs in the system. Examples include the number of tuples processed per operator, the number of tuples sent by an operator, and the number of tuple bytes processed by a PE. Custom metrics are related exclusively to the semantic of each operator type available in SPL. For example, a filter operator may maintain the number of tuples it discards. Operators can create new custom metrics at any point during their execution.

SPL allows applications to *import* and *export* streams to/from other applications. Developers must associate a stream ID or properties with a stream produced by an application, and then use such ID or properties to consume this same stream in another application. When both applications are executing, the SPL runtime automatically connects the exporter and importer operators. Importing and exporting streams enables many scenarios in which application orchestration is important, such as incremental application deployment and live application maintenance.

### 2.2 System S Runtime

The System S runtime infrastructure has three main components, namely the *Streams Application Manager* (SAM), the *Streams Resource Manager* (SRM), and the *Host Controller* (HC).

The main responsibility of the SAM daemon is to receive application submission and cancellation requests. Each application submitted to SAM is considered a new *job* in the system. When starting a job, SAM spawns all the PEs as-

1828

sociated with that application according to their placement constraints. SAM can also stop and restart PEs running in the system.

The SRM daemon is responsible for maintaining information regarding which hosts are available to the System S runtime for application deployment. It also maintains status information about which system components (e.g., SAM) and PEs are up and running. The SRM is responsible for detecting and notifying the occurrence of process or host failures. This daemon also serves as a collector for all metrics maintained by the system, such as the built-in and custom metrics of all SPL applications under execution.

The HC is a local daemon residing in each host of the system that can run SPL applications. This daemon does local operations on behalf of the central components of the system, such as starting local processes for running PEs and maintaining process status information. The HC also collects metrics from PEs running locally and periodically send them to SRM.

## 3. ORCHESTRATOR ARCHITECTURE

An orchestrator is composed of two parts. The first part is the *ORCA logic*, which contains the application-specific control code. The ORCA logic can be used to start and control *one or more* streaming applications. The second part is the *ORCA service*, which is a daemon that provides the ORCA logic a set of interfaces for event handling and an API to help the implementation of actuation routines.

Developers write the ORCA logic in C++ by inheriting an `Orchestrator` class. The `Orchestrator` class contains the signature of all event handling methods that can be specialized. The ORCA logic can invoke routines from the ORCA service by using a reference received during construction. Note that the ORCA logic can only receive events and act on applications that were started through the ORCA service. If the ORCA logic attempts to act on jobs that it did not start, the ORCA service reports a runtime error.

The result of the compilation of the ORCA logic is a shared library. We also create an XML file, which contains the basic description of the ORCA logic artifacts (e.g., ORCA name and shared library path) and a list of all applications that can be controlled from the orchestrator. Each list item contains the application name and a path to its corresponding ADL file. The ORCA service uses the ADL file to start applications and to create an in-memory stream graph representation of all the applications being managed.

To support the execution of an orchestrator, we extended System S to consider orchestration as a first class concept in its runtime infrastructure. This means that the runtime components of System S (e.g., SAM and SRM) are aware of an orchestrator as a manageable entity. For example, SAM keeps track of all orchestrators running in the system and their associated jobs.

Figure 4 shows how the System S runtime handles orchestrator instances. Similar to the process of submitting an application, users submit the orchestrator description file (MyORCA.xml) to SAM. SAM then forks a process to execute the ORCA service. Executing the orchestrator in a new process ensures memory isolation between user-written code and infrastructure components. On its startup, the ORCA service loads the ORCA logic shared library (MyORCA.so) and invokes the orchestrator start event callback. The ORCA logic can then call any functionality available from the ORCA service. As a result, the ORCA service can issue and receive calls to/from infrastructure components, such as SAM, SRM, and operators belonging to a managed application.

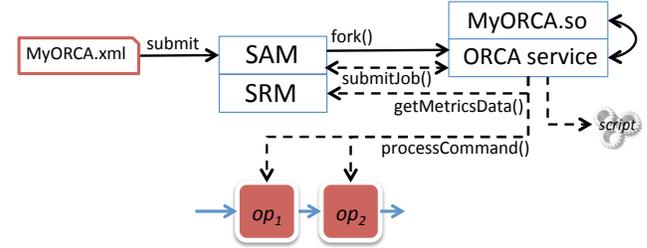

Figure 4: Users submit a new orchestrator to SAM, which instantiates a new process for running the ORCA logic. The specified logic can result in the ORCA service issuing requests to SAM, SRM, and to operators that belong to one of the managed applications.

The figure also shows the external components that the ORCA service may interact with to generate events to the ORCA logic. More specifically, the ORCA service generates component failure events once SAM *pushes* a failure notification. The generation of such an event does not add a performance penalty to the managed applications, since we are reusing the failure detection mechanisms already available in the System S infrastructure. The handling of such an event by the orchestrator, however, can increase the recovery time of the application, since the failure reaction is delayed by one extra remote procedure call (from SAM to ORCA service) plus the time consumed by the user-specific failure handling routine. The ORCA service generates runtime metric events by *pulling* such data from SRM at a specified rate. This call does not have a direct effect on application performance because getting metrics from SRM do not generate further remote calls to operators. Operators and PEs deliver updated metric values to SRM at fixed rates (every 3 seconds, by default) independent of orchestrator calls. The ORCA service can also receive user-generated events via a command tool, which generates a direct call to the ORCA service. This direct call also does not interfere with the application hot path. This interaction is not shown in the diagram for simplicity purposes. The above performance impact assessment is specific to the System S infrastructure. The impact of providing application information to an orchestrator managing another streaming platform depends on how each system detects and disseminates event data.

Note that the ORCA service and the ORCA logic provide additional functionality to the system and do not substitute the role of other infrastructure components (e.g., deployment of individual jobs). In addition, the *orchestrator does not substitute local adaptation policies that are application-independent*, which are, in general, specialized to certain operator types and that take place at the operator code itself (e.g., a dynamic filter operator that changes its filtering condition during runtime upon receiving a control command). The ORCA service acts, in fact, as a proxy to issue job submission and control commands. The ORCA logic implements an *application-specific management policy* by using the ORCA service to enforce adaptation conditions and the order in which control commands must take place.

## 4. BUILDING AN APPLICATION MANAGEMENT POLICY

To build a management policy with the orchestrator, developers must specify *which* events are of interest and *how* the application should adapt upon the occurrence of these events. With the orchestrator, this logic is specified in the ORCA logic by using the APIs provided by the ORCA service (e.g., actuation methods that are applicable to all streaming applications). In this section, we describe in more detail the services built into ORCA service that facilitate the implementation of management policies.

### 4.1 Event Scope

The ORCA service can deliver two different sets of events. The first set has events generated by the ORCA service itself. This includes the following notifications: a start signal, job submission, job cancellation, and timer expiration. The second set of events requires the ORCA service to interact with external middleware runtime components. This includes events related to application metrics, failure events, and user-defined events.

To simplify the development of the ORCA logic and reduce the number of notifications received during runtime, developers can specify the *event scope* they are interested in. In our design, the only event that is always in scope and must be handled by the ORCA logic is the *start* notification. For other events, developers must explicitly *register* with the ORCA service *event scope*. The ORCA service event scope is composed of a disjunction of *subscopes*. The ORCA service delivers an event to the ORCA logic when it matches at least one of the registered subscopes. The ORCA service delivers each event only once, even when the event matches more than one subscope.

Creating a subscope to be registered with the ORCA service requires the definition of which type of events the application control logic needs. Examples of event types include PE failures, operator metrics, PE metrics, and operator port metrics. Subscopes can be further refined based on the different attributes of an event. For example, one attribute of a metric event is its name. A subscope can define a filter on these attributes, such as asking for operator metrics that have a given metric name. Other available event attributes include application related attributes (e.g., application name) and attributes of the subgraph of the application that the event is contained within (e.g., type of the composite operator that contains the event). This fine-grained filtering is enabled by the stream graph representation maintained by the ORCA service for all applications being managed. Filtering conditions defined on the same attribute are considered disjunctive (e.g., as asking for an event that is associated with application A *or* application B), while conditions defined on different attributes are considered conjunctive (e.g., as asking for an event that is associated with application A *and* contained within composite operator type `composite1`). The ORCA logic can register multiple subscopes of the same type.

Figure 5 shows a code segment of the ORCA logic for the application in Figure 2. This ORCA logic receives events that match two different kinds of subscopes. The first subscope is of type operator metric (`OperatorMetricScope`, line 04). It matches events related to a limited set of operators with specific type (lines 05-06) and related to one specific metric (lines 07-08). Note that developers can specify subscopes by considering the application structure. For example, the invocation to `addCompositeTypeFilter` (line 05) results in only operators residing in a composite of type `composite1` being considered for event delivery. The invocation to `addOperatorTypeFilter` (line 06) leads to an additional filtering condition, which mandates only events associated with operators of type `Split` and `Merge` to be delivered. Once the ORCA service receives the `oms` subscope registration (line 13), the ORCA logic can receive operator metric events for all metrics named `queueSize` from operators of type `Split` or `Merge` that are contained in any instance of a composite operator of type `composite1` (i.e., `queueSize` metric events for operators $op_{3'}$, $op_{3''}$, $op_{6'}$, and $op_{6''}$ in Figure 2). The second subscope matches PE failure events (`PEFailureScope`, line 10). This subscope only has an application filter (`addApplicationFilter`, line 11), so failure events affecting PEs that contain any operator in application `Figure2` are delivered to the ORCA logic.

```
01: void MyOrca::handleOrcaStart(
02:   const OrcaStartContext & context)
03: {
04:   OperatorMetricScope oms("opMetricScope");
05:   oms.addCompositeTypeFilter("composite1");
06:   oms.addOperatorTypeFilter({"Split", "Merge"});
07:   oms.addOperatorMetric(
08:     OperatorMetricScope::queueSize);
09:
10:   PEFailureScope pfs("failureScope");
11:   pfs.addApplicationFilter("Figure2");
12:
13:   _orca->registerEventScope(oms);
14:   _orca->registerEventScope(pfs);
15:
16:   // other operations here
17: }
```

Figure 5: Example code for specifying and registering event scopes with the ORCA service. ORCA service delivers operator metric events for metrics named `queueSize` and associated with operators of type `Split` or `Merge` residing in a composite of type `composite1`. Any PE failure event associated with the application `Figure2` is also delivered.

The proposed API offers a much simpler interface to developers when compared to an SQL-based approach to express event scopes. This is because stream application graphs can be built by assembling operators and composite operators. Composite operators can contain other composite operators, which require the use of recursive queries to fully evaluate operator containment relationships. The SQL query below is equivalent to the `OperatorMetricScope` described in Figure 5. In this SQL query, we assume that each event attribute is represented as a table. For simplicity, we assume that composite and operator types are attributes of the composite and operator instance tables.

```
WITH CompPairs(compName, parentName) as (
  SELECT CI.compName, CI.parentName
    FROM CompositeInstances as CI
  UNION ALL
  SELECT CI.compName, CP.parentName
    FROM CompositeInstance as CI, CompPairs as CP
    WHERE CI.parentName = CP.compName)
SELECT metricValue
  FROM OperatorMetrics as OM, OperatorInstances as OI,
```

```
      CompositeInstances as CI, CompPairs as CP
WHERE
  OM.metricName = 'queueSize' and
  OM.operName = OI.operName and
  (OI.operKind = 'Split' or OI.operKind = 'Merge') and
  CI.compKind = 'composite1' and
  (OI.compName = CI.compName or
      (OI.compName = CP.compName and
       CI.compName = CP.parentName))
```

## 4.2 Event Delivery

The ORCA service delivers to the ORCA logic all events matching at least one of the registered subscopes. Events are delivered to the ORCA logic one at a time. If other events occur while an event handling routine is under execution, these events are queued by the ORCA service in the order they were received.

For each event, the ORCA service delivers two items. The first item is an array that contains the keys for all subscopes that match the delivered event. Developers associate a key with a subscope when the subscope is created (lines 04 and 11 in Figure 5). The second item is the *context* of the event. The context contains a slice of the application runtime information in which the event occurs. The context has the minimum information required to characterize each type of event. Developers can use the context to further query the ORCA service and inspect the logical and physical representation of the application. Some examples of inspection queries are: Which stream operators reside in PE with id $x$? Which composites reside in PE with id $x$? What is the enclosing composite operator instance name for operator instance $y$? What is the PE id for operator instance $y$? Using the ORCA inspection APIs together with the event context enables the developer to put the received event into perspective with respect to the logical and physical views of the application. The ORCA logic can use this information to decide the appropriate management action to execute.

For delivering built-in and custom metric-related events, the ORCA service periodically queries the SRM infrastructure component. The query frequency has a default value (15 seconds), but developers can change it at any point of the execution. Because SRM's response contains all metrics associated with a set of jobs, many of them can match the current ORCA service event scope at the same time. For each metric that matches the event scope, the ORCA service delivers one event. To facilitate the identification of metric values that are measured in the same round (i.e., pertaining to the same SRM query response), the ORCA service adds an *epoch value* to the event context. The epoch value is incremented at each SRM query and serves as a logical clock for the ORCA logic. This epoch value can be used when the event handling routine needs to evaluate if multiple metrics together meet a given condition.

Figure 6 shows a code segment for handling a subset of the events matching the operator metric scope registered in Figure 5. The routine uses the `context` object to identify which operator instance name the event is associated with (lines 05 and 10). The context also contains information about the metric name (lines 07 and 12), its value (lines 09 and 14), and epoch (lines 08 and 12). Note that the tests for metric name (lines 06-07 and 11-12) are for illustration purposes only. This is because the only metric in the ORCA event scope is `queueSize`, so events associated with any other metrics are not delivered. In line 19, the event handling routine uses the epoch values to identify if the `queueSize` metric measurement for both operators op3' and op6' occur at the same logical time.

```
01: void MyOrca::handleOperatorMetricEvent(
02:     const OperatorMetricContext & context,
03:     const std::vector<std::string> & scope)
04: {
05:     if (context.instanceName == "op3'"  &&
06:         context.metric ==
07:         toString(OperatorMetricScope::queueSize)) {
08:       _op3PrimeEpoch = context.epoch;
09:       _op3PrimeValue = context.value;
10:     } else if (context.instanceName == "op6'" &&
11:         context.metric ==
12:         toString(OperatorMetricScope::queueSize)) {
13:       _op6PrimeEpoch = context.epoch;
14:       _op6PrimeValue = context.value;
15:     } else {
16:       return;
17:     }
18:
19:     if (_op3PrimeEpoch == _op6PrimeEpoch) {
20:       // adaptation logic here
21:     }
22: }
```

**Figure 6: Example code for handling an operator metric event. ORCA service delivers the context information and all matched scopes.**

The ORCA service delivers PE failure events immediately after receiving a notification that such an event occurred from SAM. When SAM receives a PE crash notification (e.g., due to an uncaught exception), it identifies which ORCA service manages the crashed PE and then informs the ORCA service that a PE has crashed. It provides the PE id, the failure detection timestamp, and the crash reason. The ORCA service also adds an *epoch value* to the PE failure event context, which allows developers to identify that different PE failure invocations are related to the same physical event. The ORCA service increments the epoch value based on the crash reason (e.g., host failure) and the detection timestamp.

## 4.3 Application Placement and Partitioning

As described in Section 2.1, SPL allows to specify both host placement and partitioning annotations for each operator in the application. Users specify these annotations based on performance, fault tolerance, and resource requirements (e.g., operators need to run in a host that has a special hardware device). Being able to influence the application partitioning and placement configurations is critical to develop effective orchestration policies.

A *host placement* configuration indicates in which hosts operators should reside. In System S, developers specify hosts by creating host pools, which contain a list of host names or tags. Influencing host placement is useful when developing management policies that require different applications to run in different hosts. One such example is a policy that manages replicas of the same application. If two replicas run on the same host, a host failure results in the crash of both replicas, defeating the objective of the policy. The ORCA service exposes a method that changes the configuration of a given application to run only in exclusive host pools, i.e., in sets of hosts that cannot be used by any other application. When developers call this method, the ORCA service modifies the application ADL to update all its host pool configurations [11]. The host pool configuration change must occur before the application is submitted, since the host pool is interpreted by the SAM component when instantiating the PEs of the submitted application.

A *partitioning* configuration specifies which operators should be placed in the same PE (i.e., the same operating system process). One example where changing operator partitioning can be convenient is when writing policies that involve restart of operators that are logically related. If logically related groups are not isolated into multiple PEs, the restart of one group may force the restart of other logically related groups, resulting in a cascading application restart. In SPL, reconfiguring operators into different PEs require application recompilation. Our current implementation of the ORCA service does not support application recompilation, but this is trivial to implement by exposing interfaces that annotate the application source code with partitioning constraints and triggering application recompilation based on that, assuming application source code is available at runtime.

### 4.4 Application Sets and Dependencies

A common scenario in stream processing is to compose a solution based on a *set* of applications. For example, applications can consume streams or files produced by other applications. This practice has many benefits. The first is that it reduces development effort by promoting code reuse. The second is that it reduces the computational resources required to run applications that use the same data streams. This is because the reused application is instantiated only once and its output is dynamically routed to other applications that want to consume the streams it produces. The third is that it enables applications to evolve over time. For example, parts of the application can be brought up and down on demand based on runtime conditions. In SPL, applications can be dynamically composed by *importing* and *exporting* streams (Section 2.1), or by using special input adapter operators that, for example, scan a given directory on the file system.

Despite its benefits, dynamic application connection can also generate inconsistencies and waste of resources. Inconsistencies occur because users can bring applications up and down at any time. As a result, an application $A$ being used by application $B$ can suddenly be cancelled, making application $B$ starve. Another situation is when an application consumes results from another application that is still building up its state and producing inaccurate results. This occurs when the application is instantiated without specific delays or without waiting for runtime conditions to be fulfilled (e.g., processing $x$ data items). Furthermore, resource waste can occur when applications continue to run even though they are not being used by any other application.

To solve these consistency and resource consumption problems, our orchestrator design allows multiple applications to be managed in the same orchestrator instance. When writing the ORCA logic, developers can register explicit *dependency relations* between different applications with the ORCA service. Based on the dependency relations, the ORCA service automatically submits applications that are required by other applications and automatically cancels applications that are no longer in use.

To take advantage of the automatic application submission and cancellation provided by the orchestrator, developers must first create *application configurations*. An application configuration consists of the following items: (i) a string identifier, (ii) the application name, (iii) a string hash map with submission-time application parameters, (iv) a boolean indicating if the application can be automatically cancelled (i.e., the application is *garbage collectable*), and (v) a float value indicating for how long a garbage collectable application should continue to run before being automatically cancelled (called the *garbage collection timeout*). Once a configuration is created for each application that must be submitted, developers can register a unidirectional dependency between two application configurations. The ORCA service returns a registration error if the registered dependency leads to the creation of a cycle. When registering a dependency, developers can also indicate an *uptime requirement*. This requirement informs that the submission of the dependent application must be delayed by a specific number of seconds after its dependency is fulfilled (i.e., the application it depends on is submitted).

Figure 7 illustrates an example application dependency graph for an orchestrator managing six different applications. Each dashed box represents one application. The annotation in the upper left side of each box indicates the configuration parameters of the application (T for `true` and F for `false` with respect to enabling garbage collection). The dashed green arrows represent the established unidirectional dependency between applications. The arc annotations indicate the uptime requirement for each dependency.

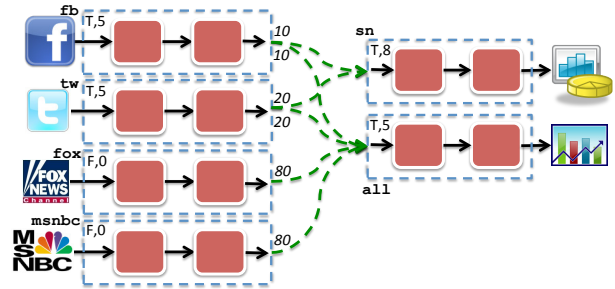

Figure 7: ORCA service internal representation of application dependencies registered in the ORCA logic. Application annotations represent configuration parameters. Arc annotations indicate how long the source application must run before the target one can start.

Once the dependencies are established, developers can submit requests for applications to start. When the start request is submitted, the ORCA service starts an *application submission thread*. This thread takes a snapshot of the current application dependency graph and cuts all nodes and edges that are not directly or indirectly connected to the submitted application (e.g., when application `all` is submitted, node `sn` and its connections to `fb` and `tw` are discarded). It then searches the graph for all applications that have no dependencies (e.g., `fb`, `tw`, `fox`, and `msnbc`), and issues start requests to SAM for all applications that are not yet running. The thread then searches for the next target application that it must instantiate and sleeps until all uptime requirements for the target application are fulfilled. The ORCA service choses an application as the next target only when all of its dependencies are satisfied (i.e., submitted) and when it has the lowest required sleeping time among all other applications with satisfied dependencies. For example, assuming that `fb`, `tw`, `fox`, and `msnbc` are all submitted at the same time, the thread sleeps for 80 seconds before submitting `all`. If `sn` was to be submitted in the same round as `all`, `sn` would be submitted first because its required sleeping time (20) is lower than `all`'s (80). The ORCA service delivers a job submission event to the ORCA logic after every application submission.

If developers issue an application cancellation request, the ORCA service automatically cancels unused applications. First, the ORCA service evaluates the application dependency graph to check if the cancellation request is issued to an application that is feeding another running application (e.g., cancellation request to `fb`). If so, the ORCA service returns an error code, enforcing that other applications do not starve. If not, it starts an *application cancellation thread* which evaluates a full snapshot of the application dependency graph to find out which applications must be cancelled. Potentially, all applications that feed the cancelled application directly or indirectly are cancelled. An application (and its dependencies) are not automatically cancelled when (i) the application is not garbage collectable (i.e., `false` is passed as a configuration in `AppConfig`, such as `fox`), (ii) the application is being used by other running applications (e.g., `fb` and `tw` feeding an instance of `sn`), or (iii) the application was explicitly submitted by the ORCA logic. The thread cancels applications following the garbage collection timeouts. These timeouts are useful when the ORCA logic submits another application that reuses an application enqueued for cancellation. This application is then immediately removed from the cancellation queue, avoiding an unnecessary application restart. For every cancelled application, the ORCA service delivers a job cancellation event.

## 5. USE CASES

This section describes three different use cases enabled by the proposed orchestrator. Each use case displays a different class of self-management policies for streaming applications. The objective of this section is to illustrate how real scenarios can benefit from our proposed framework.

### 5.1 Adaptation to Incoming Data Distribution

*Target Application.* In this scenario, we consider an application that runs a sentiment analysis algorithm on Twitter feeds, as described in Figure 1. The application first filters out all tweets that are not related to a configured product of interest. It then categorizes the tweet as containing an either positive or negative sentiment. If the tweet has a negative sentiment, it is stored on disk for later batch processing. Each tweet with a negative sentiment is then associated with a *cause*. Tweets associated with the same cause are aggregated to identify the top causes for user frustration when using the product of interest. The set of possible causes for user frustration are pre-computed using a Hadoop job [26] using IBM's Big Insights platform [13] and its module for text analytics [19]. The streaming application consumes the output of the Hadoop job during its bootup.

*Need for Adaptation.* This application must adapt to changes on the content of the incoming data. This is because the possible causes for user frustration can change over time (e.g., as companies release new product version). To adapt to these conditions, this application code contains two extra operators. The first operator consumes a stream that contains the values of two custom metrics maintained by one of the upstream operators. The metrics contain the total number of negative tweets with known and unknown causes. When the number of tweets with unknown causes is greater than the number of tweets with known causes, the operator sends a tuple to the second extra operator. The second operator then executes a script that issues a new Hadoop job that recomputes the possible user frustration causes using the file containing the latest tweets with negative sentiment. The streaming application automatically reloads the output of the Hadoop job as soon as the job finishes. By using the orchestrator, we are able to place the adaptation code purely in the ORCA logic, separating the application control and the data processing logic.

*Event Scope.* During the execution of the orchestrator start callback, we add to the scope the two custom operator metrics that maintain the total number of tweets with known and unknown causes. These two metrics values are sufficient to indicate that there are too many users complaining about an unknown issue, and, therefore, the application must adapt.

*Actuation.* When receiving operator metric events, the handling routine updates the ORCA logic private variables with the notified metric value and the metric notification epoch. If the metric epoch values for both the known and unknown causes are the same, the routine compares the metric values. If there are a greater number of tweets with unknown cause, the ORCA logic calls the script that issues the Hadoop job. The event handling routine only issues a new Hadoop job if no other job has been started in the last 10 minutes. This avoids the continuous triggering of new jobs before the application can refresh its internal state. The orchestrator has 114 lines of C++ code.

*Experiment.* Figure 8 displays the ratio of the number of tweets with unknown and known causes (y-axis) over time (or metric epoch, x-axis). In this experiment, we configured the streaming application to monitor negative sentiments related to the iPhone product. During startup, the pre-computed causes for iPhone complaints are related to Flash technology support and screen problems. At this point, the number of tweets with known causes is higher than the ones with unknown cause (ratio below 1.0). Around epoch 250, we feed a stream of tweets in which users complain about antenna issues. At this point, the graph shows the orchestrator measurement values growing, surpassing the defined actuation threshold of 1.0. Once it surpasses the threshold for the first time, the ORCA logic triggers the Hadoop job for model recomputation. Once the Hadoop job completes, the application updates its internal state. As a result, the orchestrator reports measurements below 1.0, confirming that, after the adaptation, the application can correlate the majority of the incoming tweets to a known cause.

### 5.2 Adaptation to Failures

*Target Application.* This scenario uses a financial engineering application named "Trend Calculator", which processes streams from the stock market and applies a set of financial algorithms for each incoming stock symbol over a sliding time window of 600 seconds. These algorithms include the minimum and maximum stock trade prices, the average stock price, and the Bollinger Bands above and below the average price.

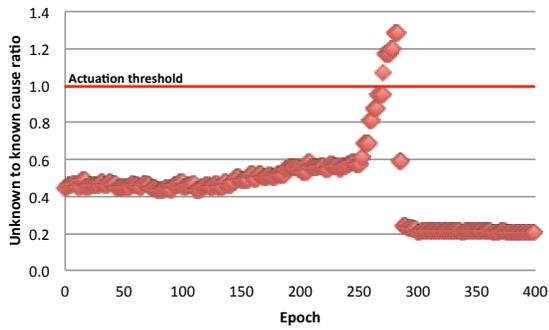

**Figure 8:** Unknown to known sentiment ratio over time (epoch). Measurement surpasses the actuation threshold defined in the **ORCA** logic, triggering the submission of the Hadoop job. Measurement stabilizes after the streaming job refreshes its model.

*Need for Adaptation.* When crash failures affect application PEs, PEs with stateful operators can lose their state and PEs with stateless operators may lead to tuple loss, which further impacts the state of the operators downstream of the failed PE [16]. To reduce end-to-end latency and increase application throughput, this application does not employ any state checkpointing feature available from the stream processing middleware. As a result, it needs to process tuples for 600 seconds to fully recover its state (i.e., until it fully refreshes its internal windows).

*Event Scope.* The only event required to provide runtime adaptation are PE failure events.

*Actuation.* To provide quick recovery for this application when PE crashes occur, we devised an orchestrator that manages three application replicas. When receiving an orchestrator start notification event, we first set the application to execute in an exclusive host pool and submit three copies of the application. The orchestrator assigns each replica an *active* or *backup* status. The status is also propagated to a file, so that a GUI can read which replica is currently active. The user of the GUI should always look for the output of the active replica for the most up-to-date results. During orchestrator startup, we also add registrations for PE failure events occurring in the target application to the **ORCA** event scope. Upon the occurrence of a PE failure and delivery of the relevant event, the **ORCA** logic evaluates if the failed replica is either active or backup. If the failed replica is the active one, then the **ORCA** logic finds the oldest running replica (i.e., replica with longest history and, most likely, with full sliding windows), sets it as active, and updates the status file. It then sets the previous active replica as backup and restarts the failed PE. The total orchestrator size is 196 lines of C++ code.

*Experiment.* Figure 9 shows the application result in two graphs that get live updates from the active replica and one of the backup replicas. The graph displays in its title the replica identifier and its current status (highlighted with solid line box), which is periodically updated based on the status file generated by the **ORCA** logic. The graph shows the results of all the financial calculations applied to a given stock of interest. Figure 9(a) shows the application results when all replicas are healthy. The orchestrator sets replica 0 as active and replica 1 as backup. When both replicas are healthy, the graphed output is identical. To forcefully trigger an orchestrator event, we kill one of the PEs belonging to the active replica. When the **ORCA** logic receives the event, it triggers an application failover and updates the status of both replicas 0 and 1. Figure 9(b) shows the graph updates after the orchestrator event handler execution. In addition to the updated status on the graph title, the figure shows the application result right after the failure occurred. The dashed line box highlights the difference between the replicas. While replica 1 continues to update the graph output, replica 0 does not produce any output while the PE is down (dashed box), and produces incorrect output right after the PE restarts and up until the application fully recovers its state.

## 5.3 On-demand Dynamic Application Composition

*Target Application.* To illustrate a dynamic composition scenario, we use an application that analyzes social media data and creates *comprehensive* user profiles. The application creates a comprehensive user profile by merging user information from multiple social media websites. For example, users with limited information in Twitter can have richer profiles in Facebook (http://www.facebook.com). The investigation of their interactions with other media sources, such as blogs and forums, can further reveal their preferences. This application is built with the following three sub-application categories:

- The first category (C1) are applications designed to extract user profile and sentiment information from social media websites that provide APIs for reading continuous streams of updates. These applications also identify some profiles based on certain criteria (e.g., profiles discussing a certain topic), which are sent out for further analysis. Our current implementation processes data from both Twitter and MySpace (http://www.myspace.com).

- The second category (C2) are applications that depend on the profiles identified by C1 applications. C2 applications query social media sources that use keyword-based search. These applications build queries based on keywords extracted from each profile. The search results are integrated into existing profiles in a data store. We currently search for extra information in Facebook, Twitter and Blogs (via BoardReader - http://boardreader.com).

- The third category (C3) are applications that use the data store produced by C2 applications to find correlations between the sentiments expressed by users and their profile attributes. We currently implement an application that correlates user sentiments with their age, gender or location.

*Need for Adaptation.* This application must *expand* itself based on the discoveries made by C1 and C2 applications. Once they find a sufficient number of profiles with new attribute fields, the application can spawn a C3 application to correlate the new attributes with user sentiments. The application must also *contract* itself once the correlation task is finished. Expanding and contracting the application over time helps decrease the overall resource consumption.

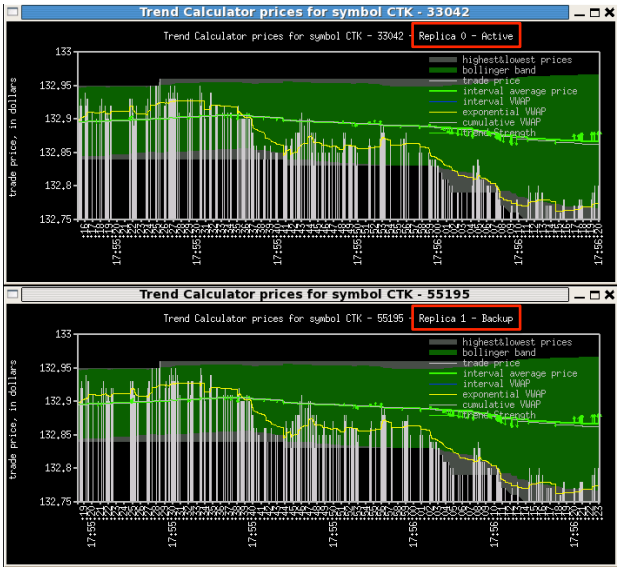
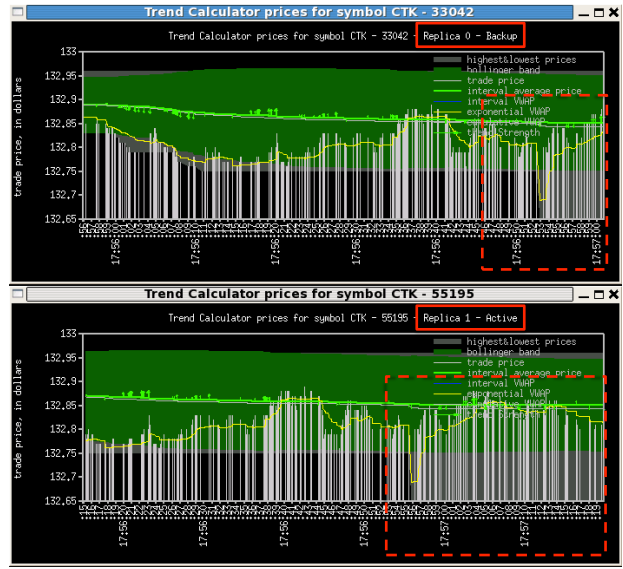

(a) Application result before PE crash  (b) Application result after PE crash

Figure 9: Figure 9(a) shows the output of the Trend Calculator application before the crash of a PE belonging to replica 0. Figure 9(b) displays the graph after the orchestrator executes the event handling routine. The dashed line box highlights the differences between the output generated by the failed and healthy replicas.

*Event Scope.* We add to the orchestrator event scope two different sets of application metrics, which are used to expand and contract the application graph on-demand. The first set is a collection of custom metrics associated to each C2 application. Each C2 application maintains a set of metrics containing the number of profiles with a specific attribute (e.g., gender). The second set is a built-in system metric of the sink operator of each C3 application. We use the final punctuation system metric to detect that the application has processed all of its tuples. In SPL, a final punctuation is a special mark in a stream that indicates that an operator will no longer produce tuples. The generation and forwarding of final punctuations are automatically managed by the SPL runtime [11].

*Actuation.* During orchestrator startup, the ORCA logic establishes dependency relations between C2 and C1 applications. The uptime requirement used for all dependencies is set to zero, since none of the C1 applications build up internal state. After the dependencies are established, the orchestrator submits all C2 applications. This results in all C1 and C2 applications to be started. When the orchestrator receives metric events associated with C2 applications, it evaluates if the aggregate number of new available profiles with a given attribute (i.e., among all C2 applications) is greater than a threshold. If so, the orchestrator submits a C3 application to compute the statistics for the given attribute. The ORCA logic evaluates the number of new available profiles based on the number of profiles available on the last C3 application submission. Note that the aggregate number of profiles may contain profiles that are duplicates. This is because C1 applications feed multiple C2 applications. Even though the orchestrator measurements include duplicates, C3 applications do not see duplicate profiles because they read directly from the data store, which has no duplicate profile entry. When the orchestrator receives a final punctuation metric event for a C3 application, it issues a job cancellation request. The orchestrator implementation has 139 lines of C++ code.

*Experiment.* Figure 10 shows the application graph visualization when applications of all categories are running in the system. In this experiment, we considered the following applications: (C1) applications consuming Twitter's 10% sample stream (`TwitterStreamReader`) and MySpace's stream (`MySpaceStreamReader`); (C2) applications searching for Twitter (`TwitterQuery`), Blogs (`BlogQuery`), and Facebook (`FacebookQuery`); and (C3) applications segmenting profiles based on age, gender, and location (`AttributeAggregator`). We also configured C1 applications to consider only profiles that issue negative posts regarding a specific product. Once the orchestrator detects that a number (e.g., 1500) of new profiles with gender, age, or location attributes were discovered, it spawns the job that does profile segmentation using the attribute of interest (e.g., gender) as a parameter.

## 6. RELATED WORK

Orchestration in the context of stream processing systems has not been studied in the literature, as very few of the existing systems support dynamic composition and external interaction with deployed application components. Existing systems often perform automatic run-time adaptation, such as adjusting the quality of the stream data based on resource availability [14], shedding load under overloaded conditions [25], and providing self-configuring high-availability [12]. While providing such management services by the runtime is often desirable, the resulting solutions are not flexible enough due to the lack of application-level knowledge within the streaming runtime. It is a significant challenge to convey an application's notion of quality, which is often defined via custom metrics, to the runtime system. Similar arguments can be made for partial fault-tolerance requirements of the applications [16], which are often hard to generalize. Our work focuses on application-driven orches-tration. We provide abstractions that enable developers to specify the appropriate orchestration behavior required by their application. A streaming system that provides a limited form of runtime adaptability is Borealis [1]. In this system, operators can adapt their behavior by receiving control signals. Our approach is more generic, since we allow other kinds of application actuations, such as dynamic application composition and invocation of external components.

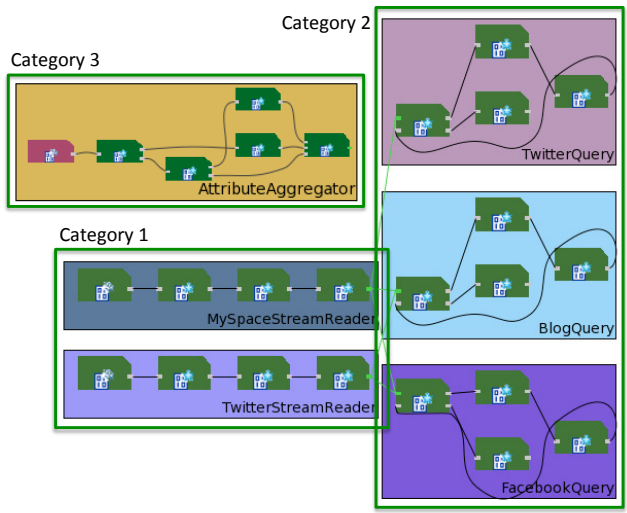

**Figure 10:** Visualization of the full application graph, which is composed of 6 sub-applications. Orchestrator expands and contracts the application graph with Category 3 application based on runtime events.

Oozie [28] is an open-source coordination service to manage data processing jobs for Apache Hadoop [26]. It orchestrates dependencies between jobs running on Hadoop. It is similar to our work in terms of the dependency management, but lacks the notion of garbage collection as Hadoop jobs are batch-based. Furthermore, it does not support any notion of subscribing to system and application-defined metrics, or introspecting the topology of the running applications.

Workflow orchestration in web services is another line of work related to ours. A web services workflow is the automation of a business process, during which documents, information or tasks are passed from one participant to another for action, according to a set of procedural rules. Business Process Execution Language (BPEL) [22] is a well adopted standard for workflow orchestration. A BPEL orchestration specifies an executable process that involves message exchanges with other systems, such that the orchestration designer controls the message exchange sequences. While similar in spirit, BPEL differs from our work due to its focus on orchestrating loosely coupled web services. As a result, it does not support registering for events of interest from the web services that are being orchestrated. It does, however, support introspecting the interfaces of the available services through their WSDL [6] descriptions.

Another area relevant to our work is orchestration within the IT infrastructure, exemplified by IBM Tivoli Orchestrator [20]. Such tools react to changes in resource requirements by taking automated provisioning actions, such as allocating servers, or installing, configuring, and patching software. Amazon's Simple Workflow Service [3] provides orchestration for cloud services. Our framework focuses on orchestration APIs for stream processing application, which can have different logical and physical deployments.

Orchestration can also be related to database tuning. Database tuning focus on acting on physical parameters of databases to optimize their performance [24, 29]. Our orchestration work focus on application specific adaptation, which can consider not only physical aspects (e.g., node placement), but application *semantics* (e.g., custom operator metrics).

Our work can also be compared to research in pervasive an ubiquitous computing, where it is common to associate a context to an event [10, 23]. For example, the context can indicate a device location change and result in application adaptation. More recently, context-aware publish-subscribe systems have been developed to provide richer information regarding message producers and consumers [7, 9]. In our orchestrator, the context also conveys information about an event. The difference is that we use the context to further query the ORCA service to inspect the application. This is critical to disambiguate the physical and logical layout of a distributed streaming application.

Finally, the aspects of our orchestration that deal with system health and unexpected errors can be compared to exception handling in programming languages. The fundamental difference lies in where the exception handling logic is executed. In our work, the orchestrator is at a central location where exceptions from multiple distributed application components can be received, and actions that require coordination across a distributed set of runtime components can be taken. This is necessitated by the distributed nature of the exception handling task [8], and is unlike programming languages where the exception handling code is embedded in a few locations within the code.

## 7. CONCLUSIONS

Due to their continuous nature, streaming applications may need to perform runtime adaptation and react to runtime events, such as failures, changes in the workload, and performance problems. In this paper, we propose the inclusion of an orchestrator component as a first class concept in a stream processing infrastructure. This allows users to (i) make their analytics and application adaptation routines more reusable by separating control and data processing logic, and (ii) to specialize event handling routines to implement adaptive stream processing applications. One of the key concepts of our orchestrator is the delivery of events with sufficient context information and an API to query graph meta-data, so that developers can disambiguate the physical and logical deployment of a streaming application. Our current orchestrator implementation works with the IBM System S infrastructure, and is able to deliver events related to failures, system-defined and application-specific metrics, and job dynamics, among others.

As future work, we plan to allow developers to dynamically add an application to the orchestrator (e.g., applications developed after orchestrator deployment). We also plan to make the orchestrator component fault-tolerant by adding transaction IDs to delivered events, and associating actuations taking place via the ORCA service to the event transaction ID. This enables reliable event delivery and actuation replay (when necessary). Furthermore, we plan to add the orchestrator concepts as an SPL ab-straction, which further facilitates the development of the ORCA logic. One option is to use rules (similar to complex event processing) for users to express event subscription more easily and take default adaptation actions when no specialization is provided for a given event (e.g., automatic PE restart).

**Acknowledgments**. We thank Kirsten Hildrum, Senthil Nathan, Edward Pring, and Chitra Venkatramani for sharing their SPL applications with us.